# Maunakea Spectroscopic Explorer (MSE): Implementing systems engineering methodology for the development of a new facility.

Kei Szeto[1a], Alexis Hill[a], Nicolas Flagey[a], Calum Hervieu[a], Mick Edgar[b], Peter Gillingham[b], Alan McConnachie[c], Shan Mignot[d], Richard Murowinski[a],

[a] CFHT Corporation, 65-1238 Mamalahoa Hwy, Kamuela, Hawaii 96743, USA
[b] Australian Astronomical Observatory, 105 Delhi Road, North Ryde NSW 2113, Australia
[c] National Research Council Canada, Herzberg Astronomy and Astrophysics, 5071 West Saanich Road, Victoria, BC, Canada, V9E 2E7
[d] GEPI, Observatoire de Paris, PSL Research University, CNRS, Univ Paris Diderot, Sorbonne Paris Cité, Place Jules Janssen, 92195 Meudon, France

## ABSTRACT

Maunakea Spectroscopic Explorer will be a 10-m class highly multiplexed survey telescope, including a segmented primary mirror and robotic fiber positioners at the prime focus. MSE will replace the Canada France Hawaii Telescope (CFHT) on the summit of Mauna Kea, Hawaii. The multiplexing includes an array of over four thousand fibers feeding banks of spectrographs several tens of meters away.

We present an overview of the requirements flow-down for MSE, from Science Requirements Document to Observatory Requirements Document. We have developed the system performance budgets, along with updating the budget architecture of our evolving project. We have also identified the links between subsystems and system budgets (and subsequently science requirements) and included system budget that are unique to MSE as a fiber-fed facility.

All of this has led to a set of Observatory Requirements that is fully consistent with the Science Requirements.

**Keywords:** requirement, system budget, SNR, noise, throughput, image quality, injection efficiency, observing efficiency

## 1. INTRODUCTION

This paper presents a systems engineering overview of the requirements flow-down for MSE, from Science Requirements Document to Observatory Requirements Document, for the Conceptual Design Phase.

In 2016, Mignot[1] presented our planned methodology for science requirements flow-down with proposed performance budgets. Since then, we have developed a set of system performance budgets meeting the science requirements and along with a budget architecture in the Observatory Architecture Document. We have also linked subsystems with science requirements through the system performance budgets, including budgets that are unique to MSE as a fiber-fed facility. All of this has resulted in a set of observatory requirements that uniquely and completely satisfies the science requirements.

Our understanding for the performance budgets has evolved into three interconnecting system budgets. In order to achieve the sensitivity requirements, we adopted the signal to noise ratio (SNR) metric from the Science Requirements Document[2] (SRD). When considering the SNR, it is clear that Throughput and Noise are obvious contributors. In addition, an Injection Efficiency budget that is unique to fiber-fed facility is utilized in order to the quantify how the target flux intensity entering the fibers at the focal surface may affect the SNR. These three budgets are interdependent and their allocations can be traded in order to optimize the SNR from bottom-up at the subsystems requirements level.

---

[1] Email: szeto@cfht.hawaii.edu; Telephone: 808-885-3188; Fax 808-885-7288

We include discussion for other relevant system budgets, including Image Quality, Point Spread Function and Observing Efficiency. Image Quality is linked to the Injection Efficiency and is defined as a separate budget for design purposes in the Maunakea Spectroscopic Explorer (MSE) paradigm. Similarly, a separate Point Spread Function budget is linked to the Noise. The Observing Efficiency budget is developed to support science operation requirements and is affected by the performance of every subsystem in MSE.

Since MSE is a spectroscopic facility, we also describe our wavelength-based calibration considerations for the sky subtraction and spectrophotometry science requirements.

## 2. SCIENCE REQUIREMENTS FLOW-DOWN

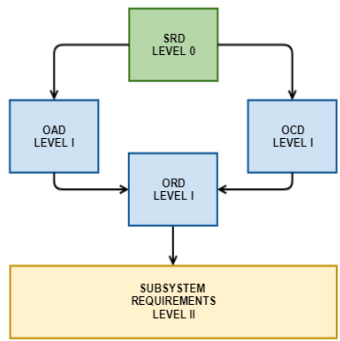

**Figure 1 SRD flow-down**

Figure 1 illustrates the systems engineering methodology we have adopted for the requirements flow-down from the MSE Science Requirements Document (SRD) to the subsystem requirements at Level 2 which trace their design requirements directly from the Observatory Requirements Document (ORD). In turn, the ORD traces its design requirements directly from the Observatory Architecture Document (OAD) and Operations Concept Document (OCD).

The SRD is considered the highest level requirements document at Level 0, and the OAD, OCD and ORD are considered next highest as Level 1 documents. The Level 0 document is developed by the scientists and under formal change control by the MSE management group. The Level 1 documents are the responsibility of the MSE Project Office (PO) and under formal change control by the MSE change control board.

This section outlines the composition of these documents and their relationship.

**2.1 Science Requirements Document**

The SRD contains 23 requirements in six groups, according to the requirement topics, Figure 2:
1. Spectral Resolution
2. Focal Plane Input
3. a. Spectral Coverage
   b. Sensitivity
4. Calibration
5. Lifetime Operations.

At the system design level, groups 1, 2, 3a and 5 lead to independent functional and operational design elements that flow-down directly into Level 1 requirements. Groups 3b and 4 however are organized into interdependent design elements in order to meet the prescribed science requirements.

The system level requirements, independent and interdependent, are captured in the OAD and OCD as shown in Figure 3. In general, the OAD prescribes engineering designs and specifies performance levels, and the OCD prescribes the technical processes and science procedures required.

| Requirements relating to Spectral Resolution: | |
|---|---|
| REQ-SRD-011 | Low spectral resolution |
| REQ-SRD-012 | Moderate spectral resolution |
| REQ-SRD-013 | High spectral resolution |
| **Requirements relating to the Focal Plane Input:** | |
| REQ-SRD-021 | Science field of view |
| REQ-SRD-022 | Multiplexing at low resolution |
| REQ-SRD-023 | Multiplexing at moderate resolution |
| REQ-SRD-024 | Multiplexing at high resolution |
| REQ-SRD-025 | Spatially resolved spectra |
| **Requirements relating to Sensitivity** | |
| REQ-SRD-031 | Spectral coverage at low resolution |
| REQ-SRD-032 | Spectral coverage at moderate resolution |
| REQ-SRD-033 | Spectral coverage at high resolution |
| REQ-SRD-034 | Sensitivity at low resolution |
| REQ-SRD-035 | Sensitivity at moderate resolution |
| REQ-SRD-036 | Sensitivity at high resolution |
| **Requirements relating to Calibration** | |
| REQ-SRD-041 | Velocities at low resolution |
| REQ-SRD-042 | Velocities at moderate resolution |
| REQ-SRD-043 | Velocities at high resolution |
| REQ-SRD-044 | Relative spectrophotometry |
| REQ-SRD-045 | Sky subtraction, continuum |
| REQ-SRD-046 | Sky subtraction, emission lines |
| **Requirements relating to Lifetime Operations** | |
| REQ-SRD-051 | Accessible sky |
| REQ-SRD-052 | Observing efficiency |
| REQ-SRD-053 | Observatory lifetime |

Figure 2 SRD requirements groups — grouped as: 1 (Spectral Resolution), 2 (Focal Plane Input), 3a (Spectral coverage), 3b (Sensitivity), 4 (Calibration), 5 (Lifetime Operations).

Figure 3 Delineation of SRD requirements into Level 1 documents — OAD – Design (Spectral Resolution, Focal Plane Input, Spectral coverage); OAD – Budget (Sensitivity); OCD – Calibration (Calibration); Lifetime Operations split: OAD - Design (REQ-SRD-051), OCD - Budget (REQ-SRD-052), OAD - Design (REQ-SRD-053).

## 2.2 Functional Analysis

To inform the OAD and OCD, we conducted a functional analysis to identity the requisite observatory activities and understand the corresponding functionalities required to obtain the science products envisaged by the MSE astronomical community.

These steps were included in our analysis and they are outlined in the following subsections:
1. Identify the end-to-end observatory activities required to produce the MSE science
2. Derive the 1st level functional blocks by organizing the activities into operational groups
3. Develop the 2nd level functional blocks architecturally representing the intrinsic functionalities within the 1st level functional blocks
4. Link the 2nd level functional blocks into logical flow processes
5. Where appropriate, develop lower level functional blocks and their flow progresses in order to inform the OAD products and OCD operations

### 2.2.1 Observatory Activities Analysis

Figure 4 shows the start-to-finish observatory activities, beginning from proposal submittal and ending with science product release in the left figure, and the same activities grouped as five functional blocks are shown in the right figure. In turns, the five functional blocks forms the 1st level elements in the MSE functional architecture.

The five 1st level functional blocks shown in Figure 5 are:
  *1. Process proposals*
This functional block implies functionalities that allow the science team to submit proposals; the time allocation committee to receive and evaluate the proposals; and a digital communication network.

  *2. Prepare surveys*
This functional block implies functionalities of an interface that enables the science team to organize and upload their survey target definitions, in the order of thousands of targets, and specify their observing conditions; a verification tool for the observatory staff to validate the "observability" of the targets; and a digital communication network.

*3. Conduct surveys*

This functional block implies functionalities of a real-time observing scheduling tool that manages concurrent survey programs; an observing database and its management tool to track the progress of the surveys for each target individually; and quick-look tool for real-time quality check of the science data during observation.

*4. Process Survey data*

This functional block implies functionalities of a set of software tools to be used for data reduction, using the best calibration information available; and a database management tool to track progress of the surveys for individual targets based on the accumulated signal to noise ratio (SNR).

*5. Closeout Survey*

This functional block implies functionalities of an archive, with access control, that mirrors the observing database and is accessible by the science team and MSE community.

There exists a sixth functional block that pertains to the continuous maintenance and support of the observatory and is not specific to any set of activities. This sixth block is shown in Figure 5. For the functional analysis, our objective is to identify the required functionalities rather than describe the physical realization of them, i.e. what is needed rather than the means to meet the needs as specified. However, lower level functionalities will inevitably inform the candidate products realized and their implementation in science and technical operations in the OAD and OCD, respectively.

**2.2.2 Functional Architecture**

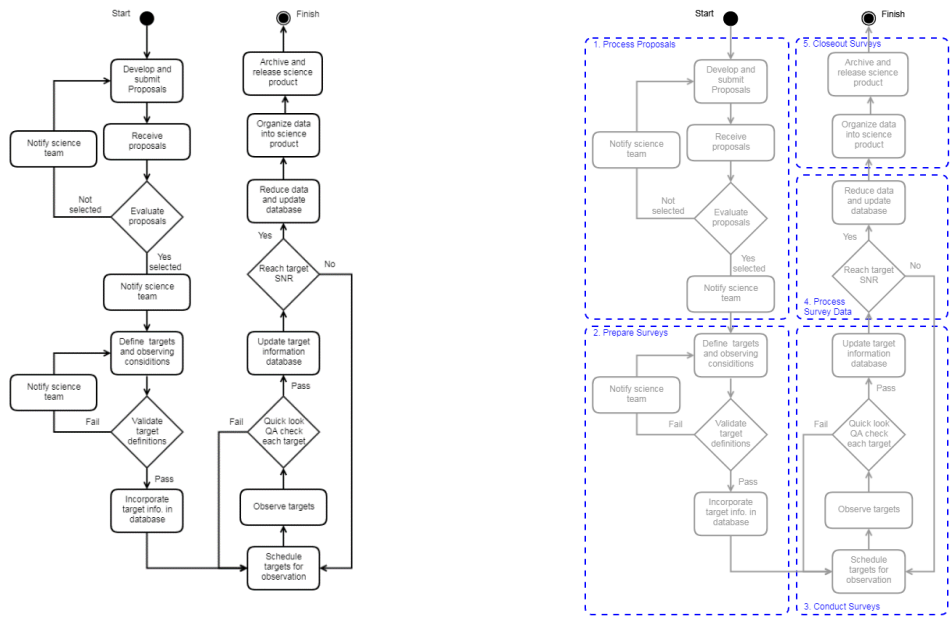

**Figure 4 Activity block diagram - observatory activities (left) and organized into five 1st level functional blocks (right).**

The functional architecture analogous to the observatory activities is shown in Figure 5 which is organized in six functional blocks described in the last section. Figure 5 lists the functionalities intrinsic to supporting the Observatory activities, not the actual products, at 2nd level functional blocks.

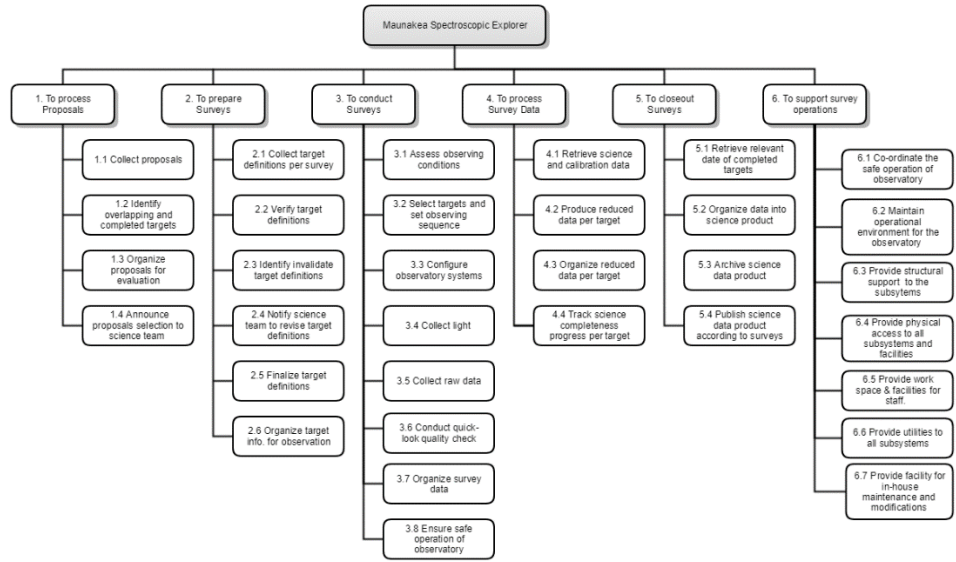

**Figure 5** Functional architecture diagram with 1st and 2nd level functional blocks

### 2.2.3 Functional Flow Block Diagrams

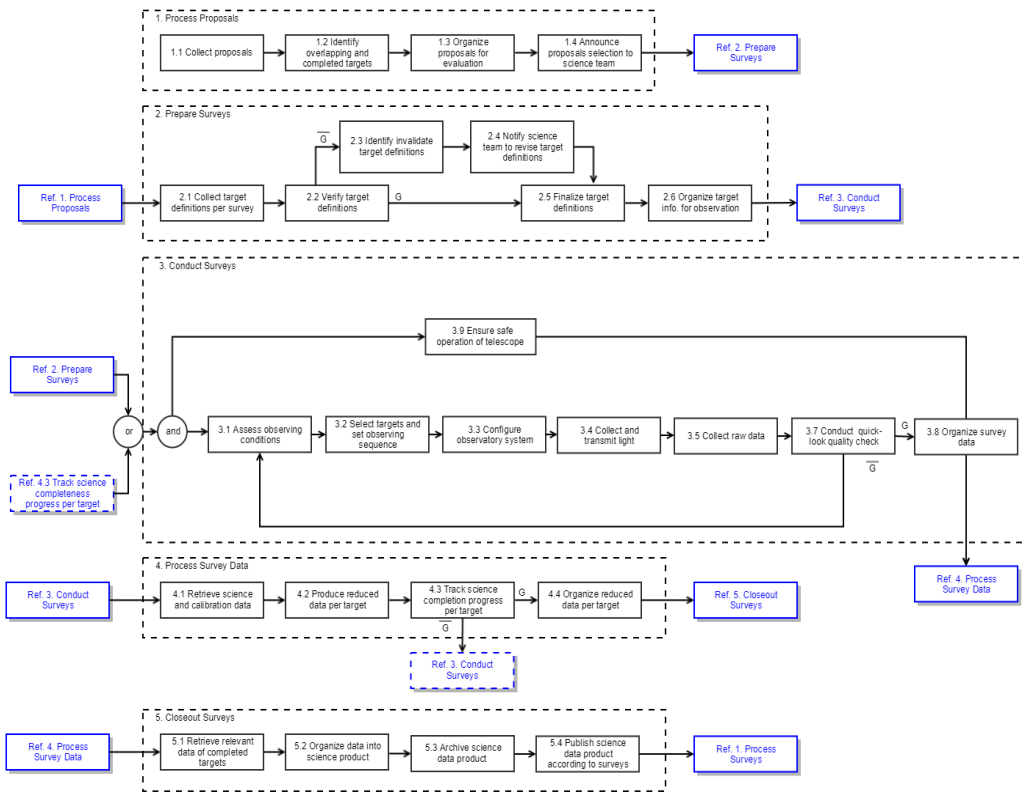

**Figure 6** Functional flow block diagram with 2nd level blocks

The corresponding functional flow block diagram for the 2nd level functions of the Observatory are shown in Figure 6 but without the sixth functional block. It is omitted as these functions must be continuously performed. Figure 6 illustrates the sequencing and interaction among 2nd level functional blocks listed in Figure 5. They represent the observatory science activities in fulfilling the MSR science.

More notable are the 2nd level functional blocks under the block *3. Conduct Surveys*. They prescribe functionalities that must be provided by physical hardware products corresponding to major subsystems within the Observatory in order to execute the planned observations, specifically within the blocks *3.3 Configure observatory system*, *3.4 Collect and transmit light* and *3.5 Collect raw data*.

Figure 7 shows the proposed hardware products corresponding to the 3rd and 4th level functional blocks under block *3.3.* The hardware identified are the enclosure, telescope mount, telescope optics, hexapod, field de-rotator, positioners and their metrology camera. Similarly, Figure 8 and Figure 9 show the proposed hardware products corresponding to the 3rd level functional blocks under block *3.4* and *3.5*, respectively. They are the acquisition and guide cameras, fiber bundles, spectrographs, calibration sources, sensors, and computer storage.

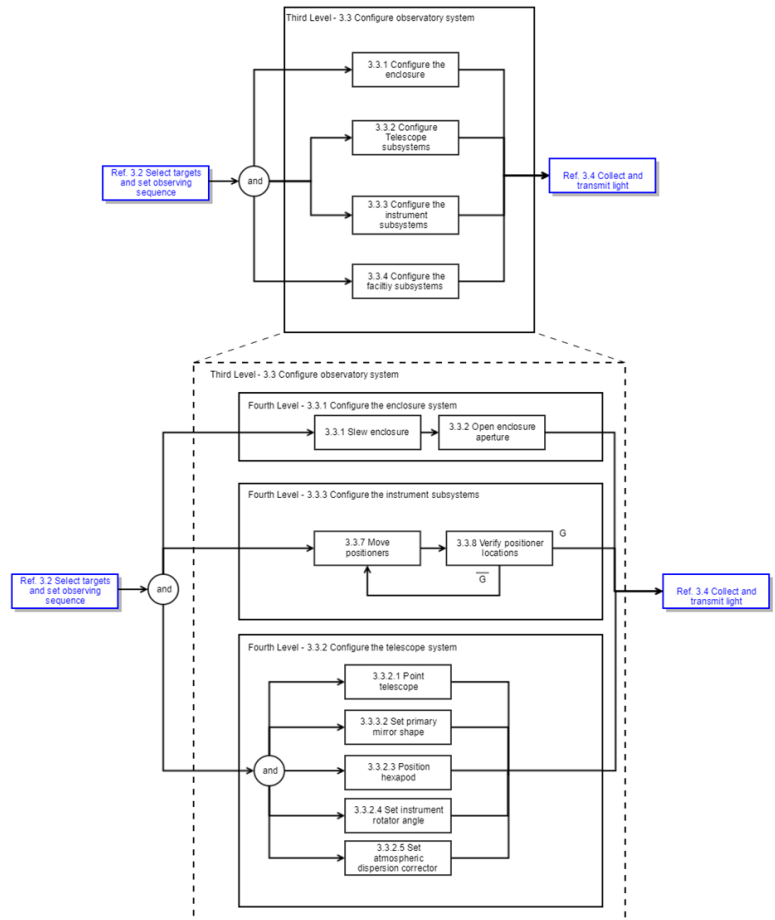

Figure 7 Functional flow block diagram of *3.3 Configure observatory system* with 3rd and 4th level blocks

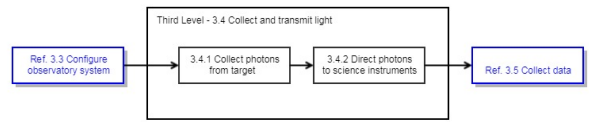

Figure 8 Functional flow block diagram of 3.4 Collect and transmit light with 3rd level functions

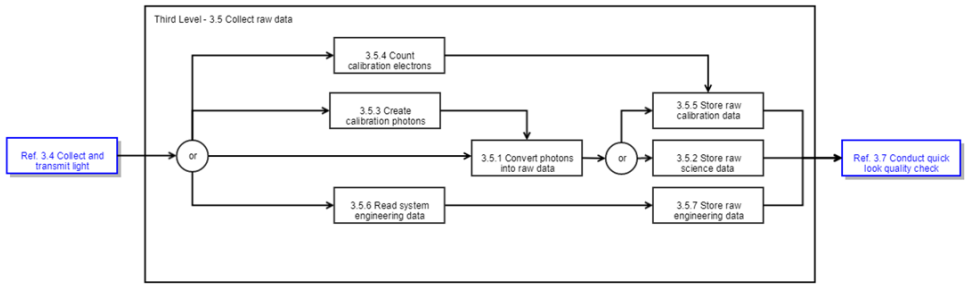

Figure 9 Functional flow block diagram of *3.5 Collect raw data* with 3rd level blocks

### 2.2.4 Product Breakdown Structure

The proposed product breakdown structure (PBS) fulfilling the functions identified in Figure 6 to Figure 9 for the MSE Observatory is shown in Figure 10. The PBS presents the design elements in a product tree structure where the Observatory is organized under six top-level products:
- Observatory Building Facilities
- Enclosure
- Telescope
- Science Instrument Package
- Observatory Execution System Architecture
- Program Execution System Architecture

#### 2.2.4.1 Observatory Building Facilities (OBF)
Structurally, the OBF contains two independent piers that support the enclosure and telescope subsystems. Operationally, the OBF contains facility infrastructure to enable science operations, including the mechanical and electrical plants, laboratories for coating of mirror segments and servicing of science instrument, shops for servicing and maintenance of enclosure and telescope components, and personnel space such as offices, technical library, staff lounge, lavatories and first aid stations, etc.

#### 2.2.4.2 Enclosure (ENCL)
The ENCL is a Calotte style dome with independently rotating base and cap structures. The base contains ventilation modules, enclosure mounted crane, telescope top-end service platform and a fixed circular shutter structure. The cap contains the aperture opening and rotates on an inclined plane at half of the telescope zenith angle atop the base structure. The combined base and cap rotations enable the enclosure to achieve the same sky coverage as the telescope.

The ENCL also includes a hardware-based safety system for protection of personnel and equipment.

#### 2.2.4.3 Telescope (TEL)
The TEL has an Alt/Az mount and it provides structural interfaces for supporting the telescope optics (segmented primary mirror and wide field corrector) and science instrument package. The telescope optics delivers a 1.5 degrees square focal surface at the telescope top-end prime focus station. The telescope structure top-end contains a hexapod system and instrument rotator system. Mechanically, the hexapod carries the wide field corrector (WFC) optical barrel and the instrument rotator, i.e. field de-rotator, which carries the prime focus station. These three subsystems are called the top-end assembly (TEA) collectively within the PO. The prime focus station contains the positioner system and two camera systems (one enables acquisition and guiding, and the other enables segment alignment, phasing and warping). Collectively, they are contained in the Telescope Optical Feedback System PBS element.

The TEL distributes the utilities required to operate its mount and control system, and top-end subsystems, and the telescope structure mounted science instrument package. The TEL also includes a hardware-based safety system for protection of personnel and equipment.

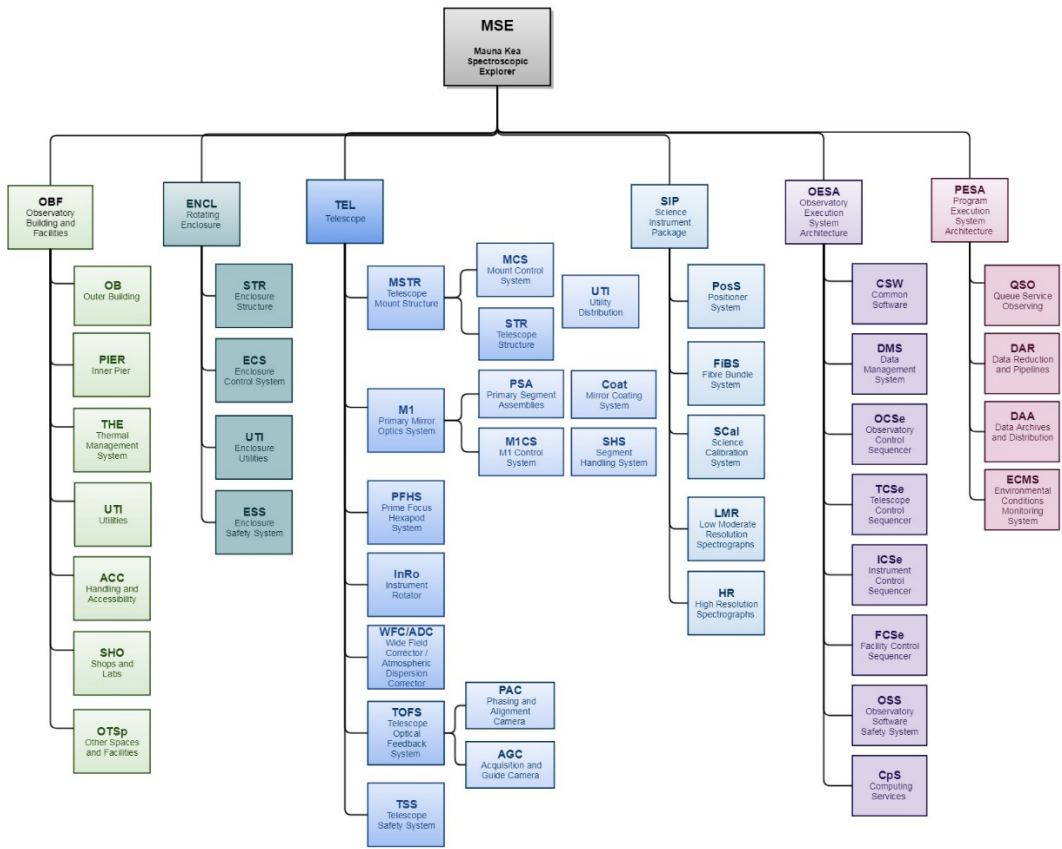

Figure 10 MSE product breakdown structure

**2.2.4.4 Science Instrument Package (SIP)**
The SIP contains the subsystems that are directly associated with obtaining science data. At the prime focus station, the positioner system for acquiring science targets by positioning the inputs of the fiber bundle system which is the light conduit between the focal surface and the spectrograph input slits. Two types of spectrographs are available, one for (switchable) low or moderate resolution (LMR) and the other for high resolution (HR), respectively. The nominal spectrograph resolutions are R3000 for LR, R6000 for MR and R40000 for HR.

The science calibration system provides the reference arcs and flats that are critical to characterize the system, end-to-end, in order to achieve the spectrophotometry and sky subtraction precision required.

**2.2.4.5 Observatory Execution System Architecture (OESA)**
The OESA contains the computer network with hardware and software control systems that enable science operations to collect and store science data as directed by the Program Execution System Architecture product described in the next section. The OESA is designed to support remote observing from the Waimea headquarters and without nighttime operator at the summit. The OESA provides an additional user interface to facilitate engineering operations for servicing and maintenance.

The OESA also includes a hardware-based global safety system for protection of personnel and equipment of the Observatory.

**2.2.4.6 Program Execution System Architecture (PESA)**
The PESA is a collection of high-level software modules that provide the functionalities to facilitate science operations, including:
- Schedule, coordinate and direct observations for science and calibration data

- Reduce and analyze the qualities of science data to provide real-time feedback for scheduling
- Archive and distribute science data, raw and reduced, along with the associated calibration and environmental information

In addition, PESA also contains sensors to monitor the environmental conditions for the purpose of scheduling, grading and monitoring the progress of science observation and ensuring the Observatory is operating within the safe environmental limits.

### 2.2.5 Product Breakdown Structure and Work Breakdown Structure

From the PBS, the corresponding Work Breakdown Structure (WBS) to realize the MSE system and subsystem architecture will be derived. The WBS relates the proposed products with the associated management, engineering, manufacturing, integration and testing tasks and the expected cost to deliver each product according to the project development phases, i.e. from conceptual design phase to final installation at the observatory site. The WBS also contains additional elements to account for the PO and the associated costs of project management, science support, engineering and administration support, etc.

In addition, the WBS assigns the interface responsibilities between PBS elements such that the scope of work for each WBS element is defined and contained. In essence, the sum of the WBS elements is a complete Observatory capable for MSE science. The purpose of the WBS is to define the scope of work and total cost of the project for management supposes.

## 3 OBSERVATORY ARCHITECTURE DOCUMENT

To summarize, the first part of OAD contains details of the aforementioned functional analysis that identifies the activities required to obtain science products outlined in the SRD, and the second part contains the aforementioned PBS representing the system architecture and its decomposition. In context of the OAD, the PBS represents the physical observatory in response to the functional analysis. The third part of the OAD specifies the corresponding system level budgets that the physical design (composed of subsystems) must meet in order to fulfill the SRD requirements, specifically performance budgets that are associated with the requirements in the SRD Group 3b. *Sensitivity*, Figure 2 and Figure 3. These performance budgets are discussed in Section 6.

Currently, the OAD contains placeholders for the point-spread function (PSF) budget and other engineering budgets, such as the mass budget, power budget, thermal budget and reliability budget, which are work-in-progress. The performance budgets have been partitioned according to physical subsystems based on information collected from their conceptual designs.

Moreover, the OAD contains high-level decisions that the PO made and may not be directly derived from the science requirements (e.g. site selection, Calotte design, prime focus telescope configuration, etc.). The system and subsystem architecture and performance budgets in the OAD are flow-down and incorporated as Level 1 requirements in the ORD.

## 4 OPERATIONS CONCEPT DOCUMENT

Based on the proposed observatory architecture in the OAD, the OCD defines an observatory organization and workflow in response to the SRD. The workflow describes the processes and procedures required to deliver the science products meeting the scientific requirements in the SRD.

The OCD considers the unique challenges of remotely operating a dedicated highly multiplexed wide-field spectroscopic survey facility and presents a detailed end-to-end operations concept for science data production. The operations concept follows the lifecycle of survey programs progressing from proposals, time allocations, target definitions, program execution, data reduction, and finally to data archive and distribution. Following the progression of parallel large surveys and PI programs, the OCD identifies the operational requirements of the observatory software tools and databases, and outlines their functionalities and utilities in order to manage survey progression and maintain quality control of the science data during different phases of operations.

Since Flagey's paper[3] in the *Observatory Operations: Strategies, Processes, and Systems* conference provides detailed description of the MSE operations concept, this section is an abbreviated description of the OCD. Readers are encouraged to reference this paper for additional information.

**4.1 Phases of Operations**

MSE follows five phases of operations customary for a ground-based astronomical facility, Figure 11. However, some phases may be executed in parallel or repeated during science operations as shown under the functional block 3. in Figure 6. Essentially the OCD phases of operations duplicate the first five functional blocks in Figure 5:

- Phase 1 (PH1): Observing program is selected
- Phase 2 (PH2): Approved targets are supplied and instrument configurations specified
- Phase 3 (PH3): Observations are executed
- Phase 4 (PH4): Data is reduced and quality analysis performed
- Phase 5 (PH5): Data products are distributed and archived

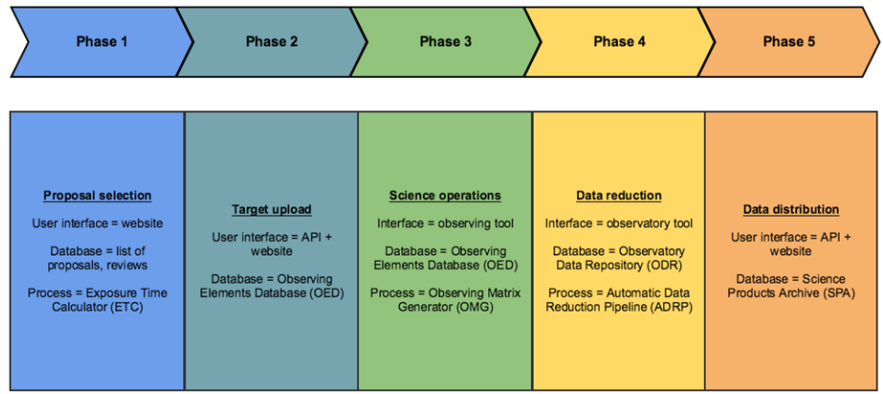

**Figure 11 Summary of the different phases of operations with software tools for processes, databases, and user interfaces.**

Figure 11 also defines the high level software tools (for processes, databases, and user interfaces) involved in these phases and the tools are discussed in details by Flagey. Their functionalities and utilities are captured in the PBS elements within the Program Execution System Architecture product in Figure 10. Specifically, the OCD describes the procedures for target selections and scheduling of parallel large surveys and PI programs utilizing the LMR and HR spectrograph sets to collect over 4,000 spectra, simultaneously, with accommodation for nighttime and daytime calibration exposures.

In addition, the OCD defines an organizational structure to support the proposed operations concept. It describes the scientific and technical staff responsibilities and skill sets required to support the science operations. The OCD also prescribes their daily workflow in ensuring the Observatory is operationally ready and outlines their interactions with the MSE community, science team and survey team.

Different from the OAD, the OCD is written narratively following the survey lifecycle progression and stating the system design requirements within each step. The design requirements are flow-down and incorporated as Level 1 requirements in the ORD.

Among the stated requirements, the OCD provides the calibration requirements associated with the SRD Group 4. *Calibration*, Figure 2 and Figure 3, and the observing efficiency requirements, based on the Observing Efficiency budget allocations, associated with SRD Group 5. *Lifetime Operations*, Figure 2 and Figure 3. They are discussed in Section 6.

# 5. OBSERVATORY REQUIREMENTS DOCUMENT

The ORD processes the OAD and OCD system level requirements and constitutes a set of cohesive Level 1 requirements. By design, the ORD serves as the "repository" where the OAD and OCD requirements are collected and interpreted into practical engineering lexicon, the same way the OAD and OCD process and interpret the SRD science requirements. In some limited cases, it is inevitable that the ORD repeats some requirements verbatim from the OAD and OCD.

The ORD is the document where all Level 2 subsystem requirements trace their origin. The intended users of the ORD are the PO staff who have intimate knowledge of the system and are tasked to prepare Level 2 subsystem design requirements documents for the Preliminary Design Phase with the subsystem design teams.

The ORD is organized into four requirement groups, namely:
1. System Constraints
2. System Requirements
3. Subsystem Requirements
4. Other System Constraints

The System Constraints group contains requirements that are externally imposed and affect all subsystems such as the site, environment, operation, regulations and standards as dictated by the Office of Maunakea Management and the County of Hawaii. Specifically, it contains requirements related to site development, observatory lifetime, operation limits based on environmental conditions, seismic events and their recovery, mass limits imposed by the OBF, and considerations for remote observing and upgradeability.

The System Requirements group contains requirements that are internally imposed by the proposed OAD system architecture and OCD operations that affect all subsystems, such as wavelength range, optical design and modes of operations. The System Requirements group also contains interdependent system budget requirements that affect relating subsystems. They are grouped together to facilitate coordination of their Level 2 requirements development. Specifically, it consolidates requirements to relevant subsystems imposed by our design choices such as the prime focus wide-field optical configuration, fiber-fed spectrograph configuration, operation plan and observing procedures, and system budgets.

The Subsystem Requirements group contains independent requirement sets for each subsystem. These requirements are organized into ten subsystem groups corresponding to the PBS.

The Other System Constraints group contains requirements that are imposed by the PO and affect all subsystems. Specifically, it contains requirements related to safety, security, environmental protection, reliability and maintainability, utilities, shipping and handling, etc.

Based on the ORD requirement groups, every Level 2 subsystem design requirements document will contain requirements that trace back to the four requirement groups.

# 6. SYSTEM BUDGETS

We adopted a system budget architecture to facilitate the flow-down of requirements for traceability purposes, i.e. budget items are expressed as design requirements and assigned to subsystems in the ORD. The major MSE system budgets are discussed in the following sections as they were presented at the System conceptual design review (Sys CoDR).

**6.1 Sensitivity Budget**

The sensitivity (SNR) budget is one of the major system budgets. It links the SRD sensitivity requirements to other system budgets and dictates many design elements of the MSE Observatory. Hierarchically, it connects five system budgets together: Noise, Throughput, and Injection Efficiency at 1st level, Image Quality and Point Spread Function (PSF) at 2nd level, as illustrated in Figure 12. Currently, the PSF function is placeholder designated for future work in the Preliminary Design Phase.

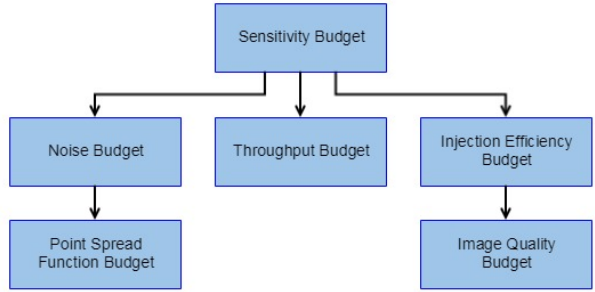

**Figure 12 System budgets hierarchical relationship**

**6.1.2 Connecting SNR with 1st Level System Budget**

McConnachie's paper[4] in this conference prescribes the set of equations to derive the wavelength-dependent SNR at each spectrograph resolution mode. Eq.1 defines SNR as the ratio of the target signal counts over the square root of the noise counts. Eq.2 lists the parameters composing the signal counts and Eq.4 shows the three contributions within the noise counts.

$$SNR = \frac{N_{obj}}{\sqrt{N_{noise}}} \quad (1)$$

$$N_{obj} = \frac{F_{obj} \times \Delta\lambda \times t \times S \times E}{P} \quad (2)$$

$$E = E_{atm} \times E_{inj} \times E_s \quad (3)$$

$$N_{noise} = N_{obj} + N_{sky} + [N_{other}]^2 \quad (4)$$

$$N_{sky} = \frac{F_{sky} \times \frac{\pi d^2}{4} \times \Delta\lambda \times t \times S \times E_s}{P} \quad (5)$$

where: N is the number of counts per resolution element
$F_{obj}$ = intrinsic flux density of target, $f(\lambda)$ — *prescribed by SRD*
$F_{sky}$ = intrinsic flux density of sky background, $f(\lambda)$ — *prescribed by SRD*
$\Delta\lambda$ = wavelength resolution, $f(\lambda)$ — *prescribed by SRD*
t = one hour observation — *prescribed by SRD*
S = telescope collection area — *prescribed by OAD*
d = input fiber diameter — *prescribed by OAD*
E = overall throughput efficiency — *prescribed by Eq.3*
$E_{atm}$ = atmospheric transmission, $f(\lambda)$ — *prescribed by OAD*
$E_{inj}$ = injection efficiency, $f(\lambda)$ — **System Budget**
$E_s$ = throughput from telescope to detector, $f(\lambda)$ — **System Budget**
$N_{noise}$ = total noise counts — *prescribed by Eq.4*
$N_{obj}$ = target counts on detector, $f(\lambda)$ — *prescribed by Eq.2*
$N_{sky}$ = sky counts on detector, $f(\lambda)$ — *prescribed by Eq.5*
$N_{other}$ = standard deviation of noise counts of other sources, $f(\lambda)$ — **System Budget**
P = energy per photon, $f(\lambda)$

Many of the parameters in Equation (1) to (5) are prescribed directly by the SRD and OAD, i.e. target flux, sky brightness, exposure time, telescope aperture, fiber diameter and atmospheric transmission at Maunakea for a given wavelength and spectrograph resolution. Only three "free" parameters remain for the SNR calculation, Equation (1), and they form the heart of the system budgets - Injection Efficiency, $E_{inj}$, Throughput, $E_s$ and Noise, $N_{other}$. However, in the SNR calculation the three parameters are not "free" but interdependent such that $N_{other}$ is linked to $E_{inj}$ and $E_s$ through the parameters $N_{obj}$ and $N_{sky}$ such that the actual value of $N_{other}$ is dependent on the values of $N_{obj}$ and $N_{sky}$. Their relationship is defined in the Noise budget in Section 6.1.4

McConnachie also outlines the strategies to balance Injection Efficiency, Throughput and Noise budgets to maximize the system sensitivity. This section provides an overview of these system budgets in the OAD. Readers are encouraged to reference his paper for detailed information regarding the methodology to maximize sensitivity adopted for the Sys CoDR.

### 6.1.3 Injection Efficiency Budget

Table 1 Injection efficiency budget

| λ, nm | 370 | 400 | 482 | 626 | 767 | 900 | 910 | 950 | 962 | 1235 | 1300 | 1500 | 1662 | 1800 |
|---|---|---|---|---|---|---|---|---|---|---|---|---|---|---|
| IE, LR % | 67.8 | 70.4 | 73.0 | 74.9 | 75.6 | 76.0 | 76.0 | 76.1 | 76.1 | 76.1 | 75.8 | 74.8 | 73.2 | 73.2 |
| IE, MR % | 67.8 | 70.4 | 73.0 | 74.9 | 75.6 | 76.0 | 76.0 | 76.1 | - | - | - | - | - | - |
| IE, HR % | 53.3 | 56.0 | 58.7 | 61.1 | 62.1 | 62.6 | - | - | - | - | - | - | - | - |

The Injection Efficiency (IE) budget is defined as the percentage of flux from a point source entering the input fiber at the telescope prime focus with respect to the total flux from the same point source corresponding to the delivered image quality and observing conditions specified in the SRD. Table 1 shows the IE budget presented at the Sys CoDR. The values were evaluated by the IE model using the subsystems conceptual design information and the anticipated system performance. The detailed modeling for the IE is presented in Flagey's paper[5] in this conference. Readers are encouraged to refer to this paper for additional information on the IE budget derivation.

Table 2 shows the subsystems and their IE budget allocations in two directions, longitudinal (z) and lateral (xy), in unit of microns according to the four budget groups. The rationales of the allocations are explained by Flagey. The IE values listed in Table 1 are calculated based on the deviations from the ideal fiber location based on the lateral and longitudinal errors listed in Table 2. The maximum amount of attainable flux would enter the fiber at z=0 and xy=0, which clearly would result in the highest Injection Efficiency value.

In Table 2, the subsystem IE allocations are partitioned and organized into four budget groups:
1. Theoretical model – as-designed system optical performance derived from Zemax
2. As-delivered – allowances based on fabrication and alignment tolerances
3. AIV – allowances resulting from the assembly, integration and verification (AIV) process
4. Operation – allowances during operation of the physical system under realistic observing conditions

**Table 2** Injection efficiency budget in longitudinal and lateral deviations.

| Budget Group | Total allocation, um | | Partition | | Discussion | WBC element |
|---|---|---|---|---|---|---|
| | Lateral | Longitudinal | | | | |
| **Theoretical Model** | 45 max | | | | Lateral chromatic aberrations | |
| | | | 45 | max | The residual chromatic aberrations after ADC correction are estimated to lead to a lateral chromatic displacement of 41 microns maximum, i.e. separation, between the foci of any two wavelengths. The separation is derived from optical design and defined by the delivered PSF computed in Zemax. | MSE.TEL.WFC/ADC |
| **As-Delivered** | | 30 max | | | Longitudinal installation errors of the combined PosS+FiTS | |
| | | | 25 | max | Scattering of PosS in Z position relative to theoretical focal surface, based on AAO-Sphinx CoDR | MSE.SIP.PosS |
| | | | 5 | max | Scattering of fibre tip distances from ferrules, based on FiTS information | MSE.SIP.FiTS |
| **Assembly Integration Verification** | | 50 max | | | Longitudinal alignment errors of the top end assembly | |
| | | | 50 | max | Residual alignment errors of the PosS+FiTS focal surface in tip/tilt and/or Z position after Assembly, Integration and Verification (AIV) | MSE.PO.ENG.AIV |
| **Operation - Focus model residual error (after setup)** | | 10 max | | | Longitudinal flexure of telescope structure due to zenith angle (gravity) and thermal changes affecting location of the fibre tips | |
| | | | 10 | max | Residual modeling error in lookup table correction for best-focus setting | MSE.PO.ENG.AIV |
| **Operation - Relative lateral fibre positioning errors (after acquisition)** | 2 max | | | | Target coordinates error | |
| | | | 2 | max | Error in target coordinates due to astrometry inaccuracy and coordinate conversion error | MSE.SCI |
| | 5 max | | | | Sky coordinates to focal surface mapping | |
| | | | 1 | max | Acquisition/guide cameras registration error with respect to the sky | MSE.SIP.TOFS |
| | | | 4 | max | Metrology system residual calibration error with respect to the focal surface | MSE.SIP.FPMS |
| | 6 rms | | | | Positioner closed-loop accuracy | |
| | | | 4 | rms | Positioner contribution based on AAO-Sphinx CoDR | MSE.SIP.PosS |
| | | | 2 | rms | Metrology system contribution based on AAO-Sphinx CoDR | MSE.SIP.FPMS |
| **Operation - Fibre defocus (spine tilt after acquisition)** | | 80 max | | | Defocus at maximum tilt for 300 mm spine with AAO Sphinx positioner (Note: defocus due to lateral positioning error is negligible, <1 um.) | |
| | | | 80 | max | | MSE.SIP.PosS |
| **Operation - Telescope motion errors (guiding during exposure)** | 5 rms | | | | Instrument rotator rotation error | |
| | | | 5 | rms | Position error resulting from imperfect control of the rotation rate; allocation of 3.5" rms rotation rate error, which corresponds to 5 um at the edge of the field of view | MSE.TEL.InRo |
| | 2 rms | | | | Based on CFHT guiding accuracy of 0.01" which corresponds to 1 um with MSE plat scale; increase allocation by 100% to 2 um; allocations add in quadrature | |
| | | | 1 | rms | Allocation for mount control | MSE.TEL.MCS |
| | | | 1 | rms | Allocation for guide camera | MSE.SIP.TOFS |
| **Operation - Differential atmospheric refraction (during exposure)** | 15 max | | | | Residual DAR drift after ADC correction | |
| | | | 15 | max | Based on optical design information and assuming observations between zenith distance 0° to 40°, where most observations are executed. | MSE.TEL.WFC/ADC |
| **Operation - Plate scale variation (during exposure)** | 10 max | | | | Plate scale variations due to unassigned optics affects | |
| | | | 10 | max | Margin to be managed by the MSE Project Office | MSE.PO.ENG.SYS |
| **Operation - Motion of fibre tips during exposure** | | 1 max | | | Gravity effect | |
| | | | 1 | max | Due to gravity sag of the positioner support structure; AAO-Sphinx CoDR reports predicted 4 um at zenith | MSE.SIP.PosS |
| | 14 max | | | | Thermal effect | |
| | | | 14 | max | Based on thermal expansion of steel focal plate, 0.59 m dia x 60 mm thk, with 2° temperature increase in one hour | MSE.SIP.PosS |
| | | 2 max | | | Thermal effect | |
| | | | 2 | max | Based on thermal expansion of steel focal plate, 0.59 m dia x 60 mm thk, with 2° temperature increase in one hour | MSE.SIP.PosS |
| | | 30 max | | | Instrument rotator axis tilt leading to longitudinal displacement | |
| | | | 30 | max | Based on 50 urad of tilt over 180° rotation in one hour | MSE.TEL.InRo |
| | | 5 max | | | Hexapod focus adjustment error | |
| | | | 5 | max | Hexapod positional errors; PFHS is assumed to adjust the focal surface position every 15 min. | MSE.TEL.PFHS |
| | 1 max | | | | Lateral fibre tip displacement due to vibration | |
| | | | 1 | max | | MSE.SIP.POsS |

**Table 3 IQ budget allocations correspond to a system IQ of 1.08" in EE80 or 0.50" FWHM**

| Item identification | | | | | Item Value | Unit |
|---|---|---|---|---|---|---|
| *Seeing* | Atmospheric | Natural site seeing | | | 0.81 | arcsec ee80 (5/3 cumulative sum) |
| | Uplift | Due to building affecting ground layer | | MSE.OBF | 0.44 | arcsec ee80 (5/3 cumulative sum) |
| | Thermal | | | | 0.22 | arcsec ee80 (5/3 cumulative sum) |
| | | Enclosure | Due to temperature gradients in dome | MSE.ENCL | | |
| | | M1 | Due to temperature difference between M1 and dome | MSE.TEL.M1 | | |
| | | Top End | Due to thermal dissipation by all components in top end | MSE.TEL.InRo | | |
| | | | | MSE.SIP.PosS | | |
| | | | | MSE.TEL.PFHS | | |
| | | | | MSE.SIP.TOFS | | |
| | | | | Seeing total | 1.02E+00 | arcsec ee80 (5/3 cumulative sum) |
| *As-Designed* | Monochromatic aberrations | | | | | |
| | WFC/ADC + idealized M1 | | | MSE.TEL.WFC/ADC | 0.25 | arcsec ee80 |
| | | | | Total design | 2.50E-01 | arcsec ee80 |
| *As-Delivered* | M1 segments figuring aberrations | | | | | |
| | Residual error after IBF | M1 | Mirror polishing and SSA mounting | MSE.TEL.M1 | 0.24 | arcsec ee80 |
| | | | | Total M1 segments | 2.40E-01 | arcsec ee80 |
| *As-Delivered* | WFC/ADC aberrations due to mounting and internal alignment errors | | | | | |
| | WFC/ADC lens figure errors | Radius of curvature | | MSE.TEL.WFC/ADC | 1.98E-02 | arcsec ee80 |
| | | Aspheric and Conic constant | | MSE.TEL.WFC/ADC | 2.17E-02 | arcsec ee80 |
| | | Slope errors | | MSE.TEL.WFC/ADC | 1.98E-02 | arcsec ee80 |
| | | Thickness errors | | MSE.TEL.WFC/ADC | 6.25E-03 | arcsec ee80 |
| | | Tilt and decentre between lens' surfaces | Relative surface-wise | MSE.TEL.WFC/ADC | 2.42E-02 | arcsec ee80 |
| | WFC/ADC materials | Homogeneity | | MSE.TEL.WFC/ADC | 1.25E-02 | arcsec ee80 |
| | Alignment errors | Tilt and decentre errors between lenses | Relative lens-wise | MSE.TEL.WFC/ADC | 1.65E-02 | arcsec ee80 |
| | | Axial separation between lenses | Barrel assembly | MSE.TEL.WFC/ADC | 6.25E-03 | arcsec ee80 |
| | | | | Total WBC/ADC as-delivered aberrations | 4.84E-02 | arcsec ee80 |
| | | | | Total As-Delivered | 2.45E-01 | arcsec ee80 |
| *AIV* | | | | | | |
| | Installed M1 segments residual error in the mirror cell | | mirror cell | MSE.TEL.M1 | 0.12 | arcsec ee80 |
| | | | | Total M1 | 1.20E-01 | arcsec ee80 |
| *AIV* | WFC/ADC barrel alignment with respect to M1 | | | | | |
| | Alignment errors | | | MSE.SIP.TOFS | 0.00E+00 | arcsec ee80 |
| | | | | Total WFC/ADC alignment | 0.00E+00 | arcsec ee80 |
| | | | | Total AIV | 1.20E-01 | arcsec ee80 |
| *Operation* | Dynamic segment alignment residuals | | | MSE.TEL.M1 | 6.20E-02 | arcsec ee80 |
| | | | | Total M1 dynamic | 6.20E-02 | arcsec ee80 |
| *Operation* | WFC/ADC barrel alignment with respect to M1 | | | | | |
| | Alignment errors | Precision of TOFS metrology errors | | MSE.SIP.TOFS | 8.84E-03 | arcsec ee80 |
| | | LUT modelling errors | | MSE.TEL.PFHS | 8.84E-03 | arcsec ee80 |
| | | | | Total WFC/ADC alignment | 1.25E-02 | arcsec ee80 |
| | | | | Total operations | 6.32E-02 | arcsec ee80 |
| | | | | **Total IQ allocation** | **1.08E+00** | **arcsec ee80** |

### 6.1.3.1 Image Quality Budget

Under the SNR budget hierarchy, the Image Quality (IQ) is intrinsically tied to the IE budget. The IQ allocations in Table 3 ensures the IQ value used in the derivation of the IE budget is met by the subsystems.

The IQ budget is defined at the 90% field radius and 30° zenith angle as per the SRD. It is expressed as the diametric 80% encircled energy (EE80) of the point spread function delivered by the "combined" telescope on the focal surface at the fiber inputs. The PSF is modeled by a 2D Moffat distribution, mathematically.

The purpose of IQ budget allocations in Table 3 is to outline the error budget in practical engineering units for the subsystem design teams. The main contributors included are seeing and telescope optics fabrication and alignment errors. The seeing includes site natural seeing, ground layer uplift and subsystem-induced thermal seeing. The telescope optics includes the M1 segments and wide field corrector with atmospheric dispersion correction (WFC/ADC). The IQ budget is organized into the same four budget groups as for the IE budget. The actual IQ allocations are based on CFHT historical data for the seeing related items, Zemax model and tolerancing results for the WFC/ADC related items, and segment performance measurements at the Keck Observatory and design budgets from the Thirty Meter Telescope project for the M1 system related items.

**6.1.4 Noise Budget**

Table 5 provides a reference summary of the Noise budget presented at the Sys CoDR for the three spectrograph resolution modes, according to the target and sky conditions specified in the SRD, Table 4. Their actual allocations are based on the subsystem conceptual designs information and the anticipated system performance. These allocations specified are consistent with the IE and Throughput budgets presented in Section 6.1.3 and 6.1.5, respectively.

**Table 4 SRD prescribed target and sky flux densities**

| Resolution | SNR/Resolution Element | | Point Source Flux Density, $F_{obj}$ | Sky Brightness, V-Band, $F_{sky}$ |
|---|---|---|---|---|
| | 370-400 nm | >400 nm | erg/sec/cm^2/Hz | mags/arcsec^2 |
| LR | ≥1 | ≥2 | 9.1xE-30 | 20.7 |
| MR | ≥1 | ≥2 | 1.4xE-29 | 20.7 |
| HR | ≥5 | ≥10 | 3.6xE-28 | 19.5 |

The Noise budget can be divided into two categories of contributors: flux-based and detector-based. The former contains sources in the system that send erroneous photons to the detectors. The latter contains error sources intrinsic to the detector, electronically and operationally. Quantitatively speaking, the flux-based portion of the Noise budget values in Table 5 are related to the IE and Throughput budgets values, and these values are only valid when used in unison.

For the set of observing conditions defined for the SRD sensitivity requirements, the science flux and sky flux are the "driving" quantities for Noise and the IE and Throughput are the "limiting" quantities for Noise. When combined they set the total flux and erroneous flux reaching the detectors. The flux-based Noise budgets are defined relatively as percentage of the combined target and/or sky flux reaching the LMR and HR spectrograph detectors. The contributions are cross-talk, ghost, diffuse light within the spectrograph and reflected from the telescope structure.

**Table 5 Noise budget, in electrons/resolution element**

| λ, nm | 370 | 400 | 482 | 626 | 767 | 900 | 910 | 950 | 962 | 1235 | 1300 | 1500 | 1662 | 1800 |
|---|---|---|---|---|---|---|---|---|---|---|---|---|---|---|
| Noise, LR | 15 | 15 | 17 | 16 | 17 | 19 | 19 | 23 | 88 | 87 | 87 | 105 | 88 | 92 |
| Noise, MR | 14 | 15 | 15 | 15 | 15 | 17 | 17 | 17 | . | . | . | . | . | . |
| Noise, HR | 16 | 17 | 18 | 17 | 17 | 17 | . | . | . | . | . | . | . | . |

The detector-based Noise budget values are functions of the design, quality and operation of the detector systems including their control electronics, and their observed thermal background. In general, they are independent of the flux level. The detector-based Noise budgets related to the electronic characteristics of the LMR and HR spectrograph detector systems are dark current and read noise.

Moreover, the Noise Budget does not consider systematic noise contributions for science calibration and data processing as they are considered separately.

## 6.1.5 Throughput Budget

The throughput requirements presented at the Sys CoDR are listed in Table 6 according to the PBS elements. The requirements are determined such that the consolidated SNR values based on the IE budget in Table 1 and Noise budget in Table 5 and Throughput in Table 6 are meeting the SRD requirements at all resolution modes.

Depending on the PBS elements, the subsystem wavelength-based throughput calculations generally include the effects of vignetting by structure and/or optics, efficiency of reflective and anti-reflection coatings, transmission and focal ratio degradation losses, and disperser efficiencies within the spectrographs.

Due to the effects of high sky absorption at ~1800nm and high sky background at ~1500nm, the corresponding SNR values are reduced drastically. As a result, the Throughput budgets in the LR mode at those wavelengths become unrealistic in order to compensate for intrinsically low SNR values. The SNR requirements will be examined in the Preliminary Design Phase by the PO to reconcile the scientific motivation and technical feasibility at those wavelengths for the LR mode.

Table 6 Throughput requirements for the PBS elements

| Wavelength (nm) | 370 | 400 | 482 | 626 | 767 | 900 | 910 | 950 | 962 | 1235 | 1300 | 1500 | 1662 | 1800 |
|---|---|---|---|---|---|---|---|---|---|---|---|---|---|---|
| ENCL | 100% | 100% | 100% | 100% | 100% | 100% | 100% | 100% | 100% | 100% | 100% | 100% | 100% | 100% |
| TEL.STR | 96% | 96% | 96% | 96% | 96% | 96% | 96% | 96% | 96% | 96% | 96% | 96% | 96% | 96% |
| TEL.M1 | 94% | 96% | 94% | 95% | 95% | 97% | 97% | 96% | 97% | 97% | 98% | 97% | 96% | 96% |
| TEL.PFHS | 99% | 99% | 99% | 99% | 99% | 99% | 99% | 99% | 99% | 99% | 99% | 99% | 99% | 99% |
| TEL.WFC/ADC | 61% | 79% | 87% | 81% | 82% | 84% | 84% | 85% | 85% | 83% | 81% | 73% | 71% | 58% |
| SIP.PosS | 97% | 97% | 97% | 97% | 97% | 97% | 97% | 97% | 97% | 97% | 97% | 97% | 97% | 97% |
| SIP.FiTS (LMR) | 46% | 58% | 72% | 82% | 85% | 86% | 86% | 84% | 85% | 79% | 79% | 74% | 79% | 61% |
| SIP.FiTS (HR) | 62% | 70% | 81% | 88% | 89% | 89% | 0% | 0% | 0% | 0% | 0% | 0% | 0% | 0% |
| SIP.LR | 18% | 11% | 21% | 18% | 17% | 23% | 24% | 44% | 45% | 41% | 43% | 167% | 61% | 261% |
| SIP.MR | 20% | 12% | 22% | 18% | 17% | 22% | 23% | 26% | 0% | 0% | 0% | 0% | 0% | 0% |
| SIP.HR | 29% | 18% | 34% | 14% | 11% | 10% | 0% | 0% | 0% | 0% | 0% | 0% | 0% | 0% |

## 6.2 Observing Efficiency Budget

The Observing Efficiency is applicable at "steady state" operations, i.e. after science commissioning, and averaged over a year. The Observing Efficiency is defined as the fraction of time the observatory is collecting photons divided by the time the observatory could have been collecting photons, which is all the time available for observations except that lost to weather.

$$Obs.Eff. = \frac{nighttime\ spent\ collecting\ photons}{all\ nighttime - nighttime\ lost\ to\ weather}$$

Flagey's paper[6] in the *Observatory Operations: Strategies, Processes, and Systems* conference provides detailed derivation of the Observing Efficiency. This section is a summary of his work and readers are encouraged to reference this paper for additional information.

Using the CFHT site information and historical weather data, he estimated the average night has 8 hours available for photon collection with observations starting after 12° twilight and developed a budget meeting the SRD Observing Efficiency requirement of 80%. After weather, the observing time loss, i.e. not collecting science photons, can be separated into two groups. The groups contains contributors from maintenance overhead and observing overhead, respectively.

The maintenance overhead contributors are losses due to primary mirror realignment and phasing after segment exchanges, failure of subsystems, and on-sky engineering time for system performance tuning. Their individual allocations are:

- on-sky engineering: **50 hours per year**
- primary mirror phasing: **21 hours per year**
- PosS failures: **30 hours per year**
- FPMS failures: **30 hours per year**
- FiTS failures: **30 hours per year**
- LMR failures: **6.67 hours per year**
- HR failures: **3.33 hours per year**
- TEL.STR failures: **5 hours per year**
- ENCL failures: **5 hours per year**
- M1 failures: **1 hour per year**
- WFC/ADC failures: **1 hour per year**
- PFHS failures: **1 hour per year**
- InRo failures: **5 hours per year**
- OBF failures: **10 hours per year**
- OESA failures: **10 hours per year**
- PESA failures: **20 hours per year**
- TOFS failures: **10 hours per year**
- SCal failures: **1 hour per year**

The total maintenance overhead loss is 240 hours per years with 69 hours allocated for subsystem failures. The budget values are based on historical data of similar systems, operation experience, subsystem conceptual designs information and engineering judgement.

For the observing overhead loss, he divided the available hours into procedural time blocks corresponding to the anticipated steps during nighttime operations outlined in the OCD. The blocks are arranged to optimize Observing Efficiency with some executed sequentially while others in parallel overlapping in time. The time blocks are defined as follows:
1. **SCI OBS** - Science observation for collecting science photons but not including the last detector readout[2]. This is the time adjustable block set to meet the Observing Efficiency requirement.
2. **SCI READ** - Last science readout at the end of SCI OBS.
3. **FPMS** - Fiber position measurements after SCI OBS.
3. **CAL CONF**, **CAL OBS**, **CAL READ** - Three sequential blocks associated with calibration data, i.e. collection of calibration photons, including source configuration, calibration exposure with sources on, and detector readout with sources off.
4. **SYS CONF** - System configuration for a new observing field. Many subsystems are setup in parallel, initially. Once pointing is achieved then the sequential process of acquisition and guiding begins.
5. **SCI CONF** - Science configuration to confirm guiding, usually after the calibration data sequence.
6. **ADRP/OMG** - Real-time process of automatic data reduction and observing queue update during observation.
7. **OVER** - Time allowance for remote observer to override the ADRP/OMG update.

### 6.2.1 Calibration Procedure

Calibrations require specific attention because they directly impact several science requirements: velocity accuracy, relative spectrophotometry, sky subtraction, and overall sensitivity. The planned procedures for extracting science-ready spectra from a fiber fed spectrograph is described by McConnachie[7] in the *Observatory Operations: Strategies, Processes, and Systems* conference. In order to provide high quality science calibration, it is essential to understand the wavelength-dependent transmission of the Observatory, i.e. the system transfer function for astronomical targets positioned anywhere in field of view, in addition to the wavelength solution and the associated PSF. For every target, we will need to know the throughput, wavelength solution and line spread function (LSF) of the spectral lines as functions of wavelength, telescope pointing, target position in the field, position and tilt of the positioner, observing time, and environmental conditions.

The MSE science calibration plan for obtaining calibration exposures is based on three principles:
1. It shall not introduce any significant sources of noise either directly as a spectral flat or indirectly through in the flux response model across the detector focal plane.
2. It shall not add significant overhead to science observations.
3. It shall be obtained in the same configuration and conditions, as close as possible, to the corresponding science observation.

---

[2] Due to the multi-wavelength arm LMR and HR spectrograph designs, exposure time varies between arms. The arm with the longest exposure time determines the duration of the SCI OBS while other arms have multiple exposures and readouts. In general, blue arm exposures are the longest and NIR arm exposures are the shortest.

In addition to the routine biases and dark frames, and pixel flats, the calibration procedure requires a sequence of high SNR "lamp" flats and arcs exposures immediately before and immediately after each science observation. This ensures the calibration exposures are obtained under system configuration and environmental conditions as close as possible to the science exposure. The lamps will illuminate the focal surface with a repeatable illumination pattern and spectral energy distribution for every fiber at a level not compromising the velocity, spectrophotometric and sky-subtraction precisions required in the SRD. For the same reason, it is essential that calibration light arrives that spectrographs mimicking the light arriving from the sky, in near-field and far-field.

The calibration procedure also requires obtaining as many twilight flats at the beginning and end of the night as possible. By taking the median[3] of numerous twilight flats ensures that all fibers have an even illumination to a high degree of certainty. We believe lamp flats, or even dome flats, do not provide the level of uniformity as twilight flats. Through the combinations of twilight flats and lamp flats taken at twilight, beginning and end, and observations, before and after, the system transfer function can be defined. Similarly, additional high SNR dome arcs are required for daytime observations and taken daily. As mentioned, dome arcs with similar far field pattern as the science light are needed such that the effective LSF is not altered. We believe dome arcs will provide the necessary corrections required by the lamp arcs in enhancing the wavelength solution.

We are cognizant of the calibration challenges in hardware design and reduction techniques. Science and engineering development to understand and resolve the calibration concerns will be the emphases of the Preliminary Design Phase.

### 6.2.2 Optimal Science Observation (SCI OBS) Duration

Incorporating the envisaged calibration procedure, Flagey examined different combinations of operation sequence to determine the lowest observation overhead, independent of the duration of the SCI OBS time block. Figure 13 shows the optimal sequence with an overhead of 271.5 seconds. He then determined the minimum SCI OBS duration required is 44 minutes in order to meet the 80% Observing Efficiency SRD requirement.

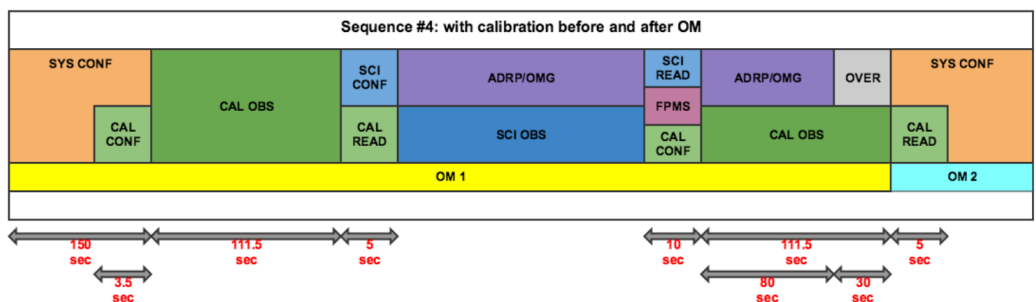

Figure 13 Optimal nighttime observing sequence

## 7. SUMMARY

For the MSE project, we have adopted a systems engineering methodology for requirements flow-down and traceability from the Science Requirements Document to the Observatory Requirements Document using two parallel and intermediary documents: Observatory Architecture Document and Operations Concept Document.

With this systems engineering methodology:
- All SRD requirements are flow-down to either the OAD or OCD.
- Non ARD requirements such as external constraints, architectural and operational considerations are introduced through the OAD and OCD and incorporated in the ORD
- All ORD requirements are traceable to either the OAD or OCD.
- The ORD is the single source where all subsystems requirements are traced.

By using the intermediate documents, this allows us to effectively convert and interpret the "abstract" scientific requirements into practical functional and operational requirements into the OAD and OCD. Through the OAD and OCD, they enable the incorporation of additional requirements from the proposed observatory architecture and operating procedures developed during the Conceptual Design Phase. This includes the system level performance budgets described. Functionally, the ORD become the clearinghouse of the OAD and OCD requirements where they are processed, consolidated and organized at the system and subsystem level to facilitate flow-down.

---

[3] This will correct for the intrinsic but predictable non-uniformity in the twilight sky across the field.

To ensure the OAD and OCD requirements are not spurious, we have performed a functional analysis to outline the observatory activities and identify the functional blocks required. By matching them with the OAD architecture in terms of the Product Breakdown Structure and with the OCD operations in terms of Phases of Operations, we can "authenticate" the OAD and OCD requirements. Conversely, the functional analysis verifies the MSE Observatory as defined by the OAD and OCD can achieve the MSE science envisaged from the perspective of a physical and operational observatory facility.

We also developed for the System conceptual design review a system budget architecture to demonstrate quantitatively the SRD sensitivity, calibration and observing efficiency requirements can be met. Currently, the sensitivity requirement is distributed among the system budgets of Signal to Noise Ratio, Through, Injection Efficiency, Noise, Image Quality and Point Spread Function. The IE budget development described is unique for a fiber-fed facility. The estimated Sys CoRD sensitivity shows deficiency for the low resolution mode and we plan reconcile the science motivation with the physical reality of atmospheric conditions.

For the observing efficiency requirement, our analysis show an 80% efficiency can be achieved by science observations that are no shorter than 44 minutes based on our analysis with science calibration as an integral part. We described the nighttime and daytime calibration plan envisaged to achieve the velocity, sky subtraction and relative spectrophotometry requirements. We also acknowledge to achieve science calibration precision will require additional engineering and scientific development in hardware design, reduction techniques and calibration procedures.

From the systems engineering perspective, the MSE project has completed the Conceptual Design Phase culminating in a successful Sys CoDR. For the Preliminary Design Phase, we plan to continue to apply our systems engineering methodology and system budget architecture in our requirements development for the Level 2 subsystem design requirements documents.

## ACKNOWLEDGEMENTS

The Maunakea Spectroscopic Explorer conceptual design phase was conducted by the MSE Project Office, which is hosted by the Canada-France-Hawaii Telescope. MSE partner organizations in Canada, France, Hawaii, Australia, China, India, and Spain all contributed to the conceptual design. The authors and the MSE collaboration recognize and acknowledge the cultural importance of the summit of Maunakea to a broad cross section of the Native Hawaiian community.